\documentstyle[11pt,aaspp4]{article}

\newcommand{\kms}{km~s$^{-1}$}
\newcommand{\kmsM} {km~s$^{-1}$~Mpc$^{-1}$}
\newcommand{\etal}{{\it et al.\/}}

\begin{document}

\title{Clustering in the Caltech Faint Galaxy Redshift Survey}
\author{Judith Cohen}
\affil{Palomar Observatory, California Institute of Technology}

\begin{abstract}

The Caltech Faint Galaxy Redshift Survey has collected $\sim$2000
spectra taken with multi-slit masks using the
Low Resolution Imaging Spectrograph at the Keck Observatory
(Oke \etal\ 1995)
in two widely separated fields on the sky,
each $\sim$1 x 1 deg$^2$.
Most of these objects are faint field galaxies; about 10\% are
Galactic stars and about 1\% are broad-lined AGNs.

I summarize the small scale clustering of this sample as
well as the evidence in support of large scale structure. 

\end{abstract}


\section{Introduction}

The defining features of the Caltech Faint Galaxy Redshift Survey 
that distinguish it from existing 
and ongoing or planned surveys
are that it is possible to reach to fainter magnitudes
($K \sim 20$~mag, $R \sim 24$~mag) and that completeness
is emphasized rather than sparse sampling over a large field. 
Objects
are observed irrespective of morphology so as not to exclude
AGNs or other unusual extragalactic sources.

The CFGRS is working on two fields.  The first field at J005325+1234.
The central survey
field measures $2 \times 7.3$~arcmin$^2$ with a statistical 
sample containing 195 
infrared-selected objects complete to $K = 20$~mag.  After
several seasons of observing, a redshift
completeness of 84\% was achieved.  
There are 139 galaxies with redshifts
as well as 24 spectroscopically confirmed Galactic stars 
and 32 objects for which spectroscopic redshifts cannot be assigned
(most of these are EROs) 
in the $K \le 20$ sample.  
There are 13 additional objects with redshifts (including two more stars)
that are just outside the field 
boundary or are within the field but too faint.
Thus in this field there are 150 galaxies with redshifts which lie
in the range [0.173, 1.44].

In addition, six-color broad band photometry from 0.36 to 2.2$\mu$
is available for
all these objects.  For the set of galaxies with redshifts,
rest frame spectral energy distributions can be derived
(assuming a cosmological model)
which reach far into the UV for the higher redshift objects.

A suite of papers containing
all the material for the field at J005325+1234
and an analysis thereof
has been published (Cohen \etal\ 1999a, Cohen \etal\ 1999b, 
Pahre \etal\ 1999).

Our second field is the HDF-North (Williams \etal\ 1996).  
Here, as described by Cohen \etal\ (2000),
after several years of effort, the spectroscopy is 92\% complete
to $R = 24$ in the HDF itself, and 92\% complete
to $R = 23$ in the Flanking Fields
covering a circle with diameter 8 arcmin centered on the HDF.  Cohen \etal\ (2000) contains a large collection of previously
unpublished redshifts as well as a compilation of all published data
for the region of the HDF; see Cohen \etal\ (2000) for details.
The total
sample of objects with redshifts in the region of the HDF now exceeds
660. 

Accompanying the spectroscopic paper is a four-color photometric
catalog of the region of the HDF with high quality astrometric
positions, Hogg \etal\ (2000). 
The $R$ band of this catalog was used
to define the spectroscopic sample completeness and the spectroscopic
sample has been matched onto this $R$-band catalog.

To facilitate work on clustering at a scale larger than that permitted
by the small solid angle subtended by the two main fields,
each of the two main fields is surrounded by outrigger fields
separated by up to 30 arcmin from the main field.  Redshifts exist for
about 500 galaxies in the outrigger fields around J005325+1234
and about 300 galaxies in the outrigger fields 30 arcmin N, S, E and
W of the HDF.

We now focus exclusively on the issue of galaxy clustering in these
samples.

\section{Evidence for the Presence of Groups to $z \sim 1.2$}

There is clear evidence for the presence of bound groups of galaxies out
to $z \sim 1.2$.  The best place to see this is in the HDF sample.  We use
a Gaussian kernel smoothing to define groups.  A $\sigma$ of 300 \kms\ is used
for the signal and a much larger smoothing width is used to define 
the mean distribution
in $z$ of the sample.  The ratio of the two functions is the overdensity
Figure 1 shows the overdensity of the HDF sample as a function of $z$ for
$z < 1.25$.  The membership and velocity dispersions of the groups
are calculated as well; see Cohen \etal\ (2000) for details.  Every
field that we have looked at, both the main fields and the outrigger fields,
shows similar small scale structure.

%
%   Figure 1
%
\begin{center}
\epsscale{0.9}
\plotone{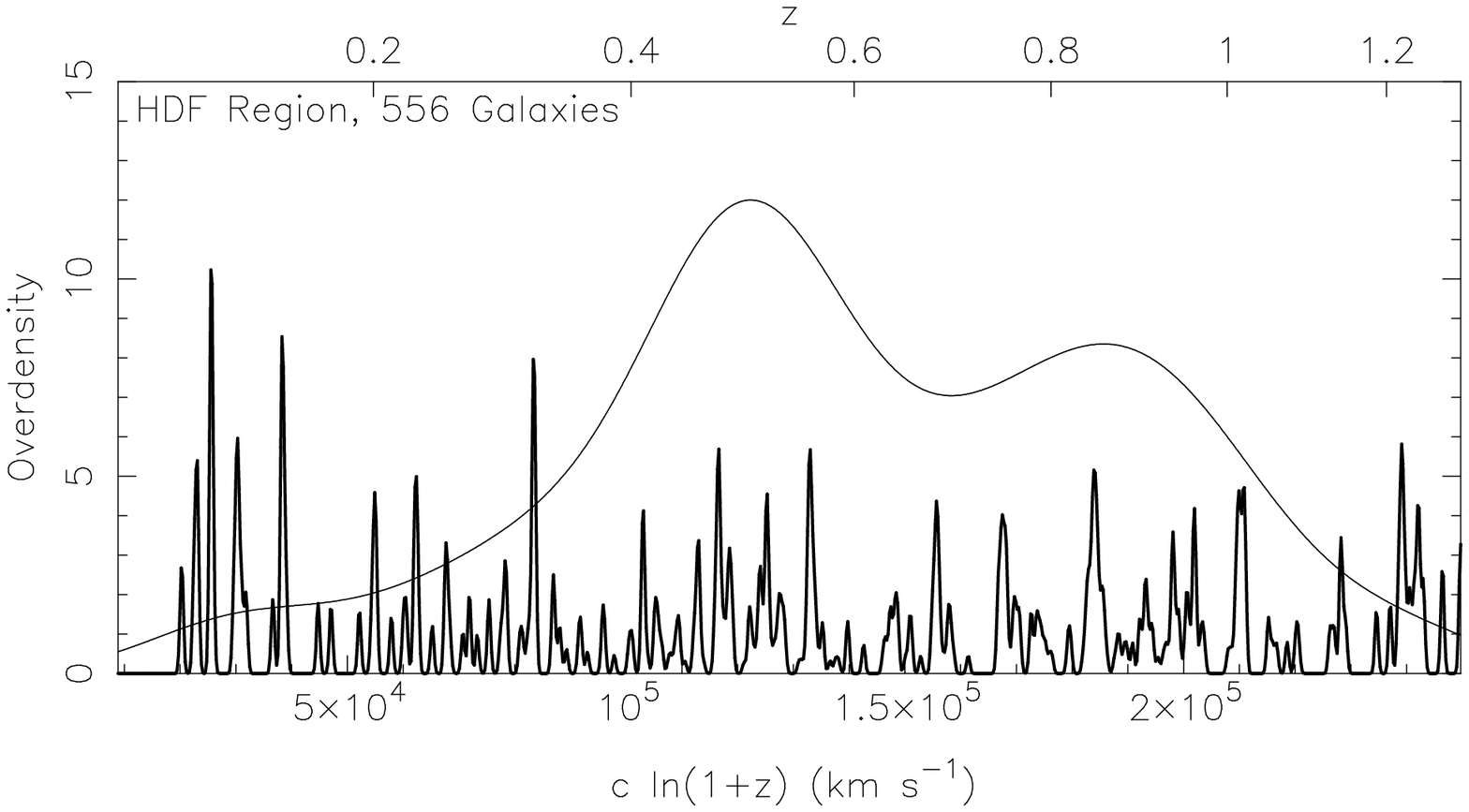} \\
\end{center}
{\sl The overdensity is shown as a function of $z$ for the HDF sample.  The
thin line is the heavily smoothed distribution of galaxies.}

\section{A Formal Analysis of Small Scale Clustering}

David Hogg has carried out an analysis of the two point correlation function
from the CFGRS database as it existed roughly 6 months ago. His results
are summarized in Figure 2, which shows the correlation length deduced
from our work for the 1 x 1 deg$^2$ area of the
field at J005325+1234 and for the 8 arcmin diameter field of the region
of the HDF.
The results are shown for three bins in $z$, and compared
to those of several recent surveys.  

This figure supports and extends earlier work which suggested that the
correlation length decreases with $z$.  There are several important
caveats.  There appear to be substantial field-to-field variations,
particularly in the lower $z$ bins, with the HDF showing abnormally
low correlation lengths.  Perhaps this has occurred because the HDF
was selected to be devoid of bright galaxies.  Also 
galaxies whose spectra are dominated by absorption lines are more
strongly clustered than are galaxies with emission lines.

%
%  Figure 2 from David HOgg
%
\begin{center}
\epsscale{0.7}
\plotone{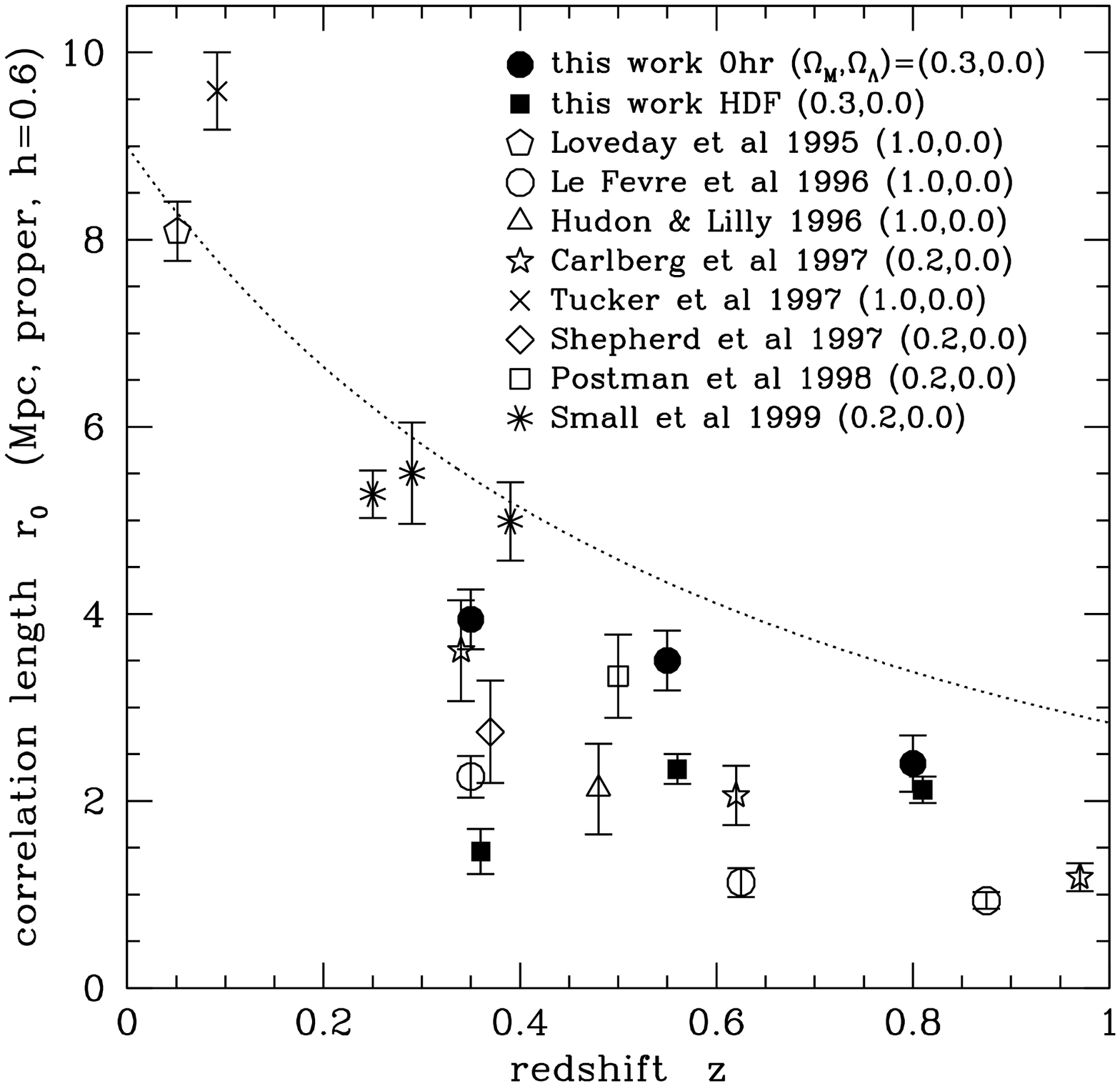} \\
\end{center}
{\sl Figure 2.
The correlation length is shown as a function of $z$ for the HDF sample
and for the 1x1 deg$^2$ sample in the field  J005325+1234.  Published
measurements for $r_0$ are all converted to H$_0$ = 60 \kmsM.}

\section{Evidence for the Presence of Large Scale Structure}

There are two ways to approach this issue using our data.  
Given the very large sample in the HDF we can assume each group/small cluster
represents the intersection of a ``wall'' similar to those seen locally
(e.g. de Lapparent, Geller \& Huchra 1986)
with the line of sight to the HDF.  One then examines the distance
in comoving coordinates between adjacent redshift peaks in the
statistically complete sample of groups to obtain Figure 3.
This figure (from Cohen \etal\ 2000) 
shows that a characteristic separation of about 70 Mpc
in our cosmology (H$_0$ = 60 \kmsM, $\Omega_M = 0.3$,
$\Lambda = 0$) seems reasonable.

%
%  Figure 3 - HDF wall separation
%
\begin{center}
\epsscale{0.5}
\plotone{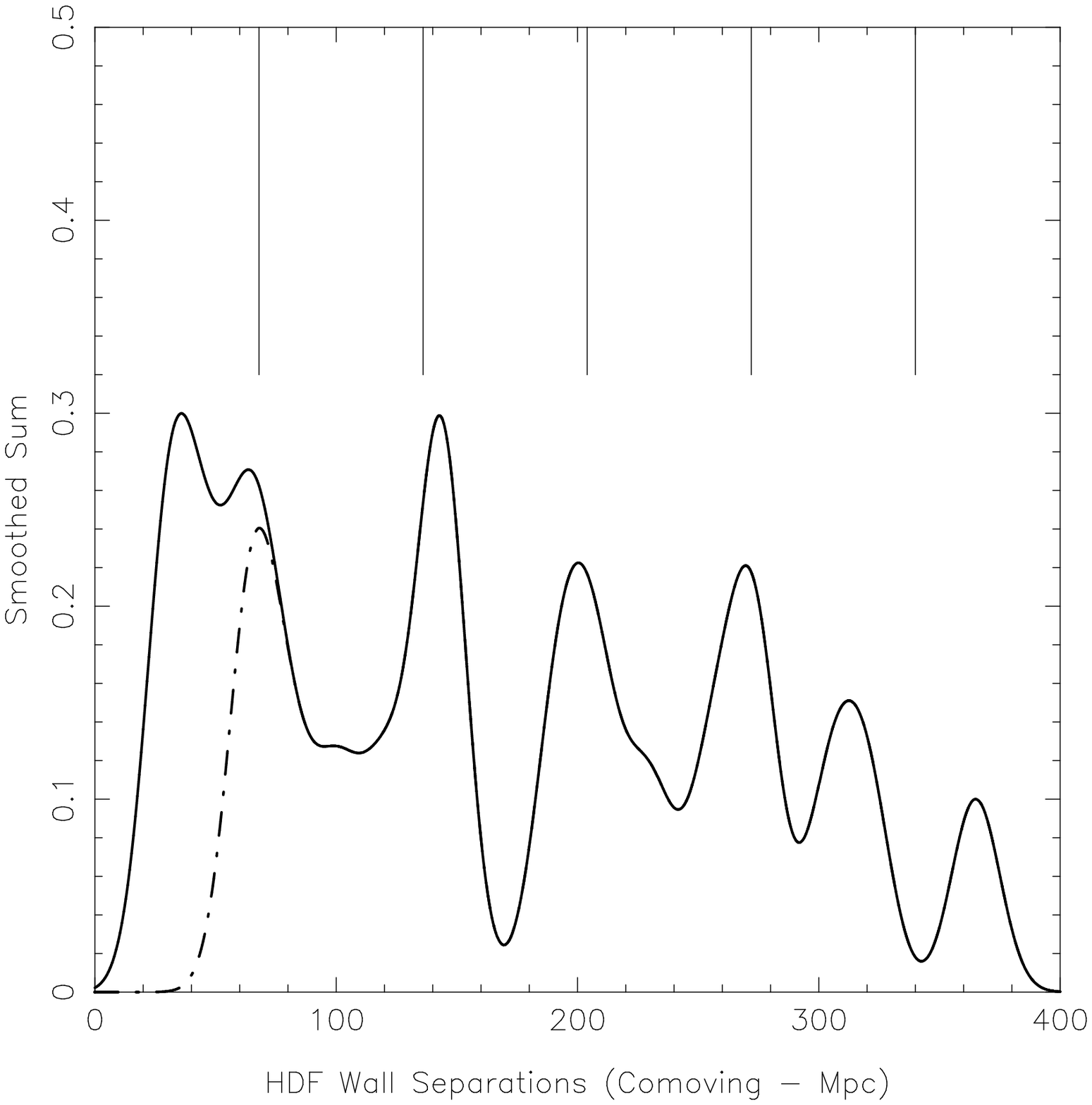} \\
\end{center}
{\sl Figure 3.  The difference in comoving distance along the line of sight
between adjacent redshift peaks 
is shown for the HDF sample smoothed by a Gaussian with $\sigma$ = 10 Mpc.
If differences less than 50 Mpc are omitted, the dashed line is obtained.
The vertical lines have a spacing of 68 Mpc. (H$_0$ = 60 \kmsM.)}

The second approach is look for correlations between the main field
and the outrigger fields across the full 1 x 1 deg$^2$ area on the sky.
While such attempts are still preliminary, the results are tantalizing.
First one needs to establish what tolerance is allowed in $\Delta(z)$ 
across this field.  This is discussed in Cohen \etal\ (2000); it is about
0.012 for $z \sim 1$.  Figure 4 shows a comparison of the
HDF with the sum of the outrigger fields around the HDF.  
It provides tantalizing evidence
for partial coherence across a scale of 1 degree.  A lot more work,
and of course a bigger sample of objects with redshifts, is going
to be required to produce a definitive result.

David Hogg and Roger Blandford are working on
a formal analysis of this data.  The non-continuous sample fields
and the selection effects for the sample within each of the fields
make this quite difficult.

%
%  Figure 4 - peaks in HDF compared to summed outrigger fields 
%
\begin{center}
\epsscale{0.7}
\plotone{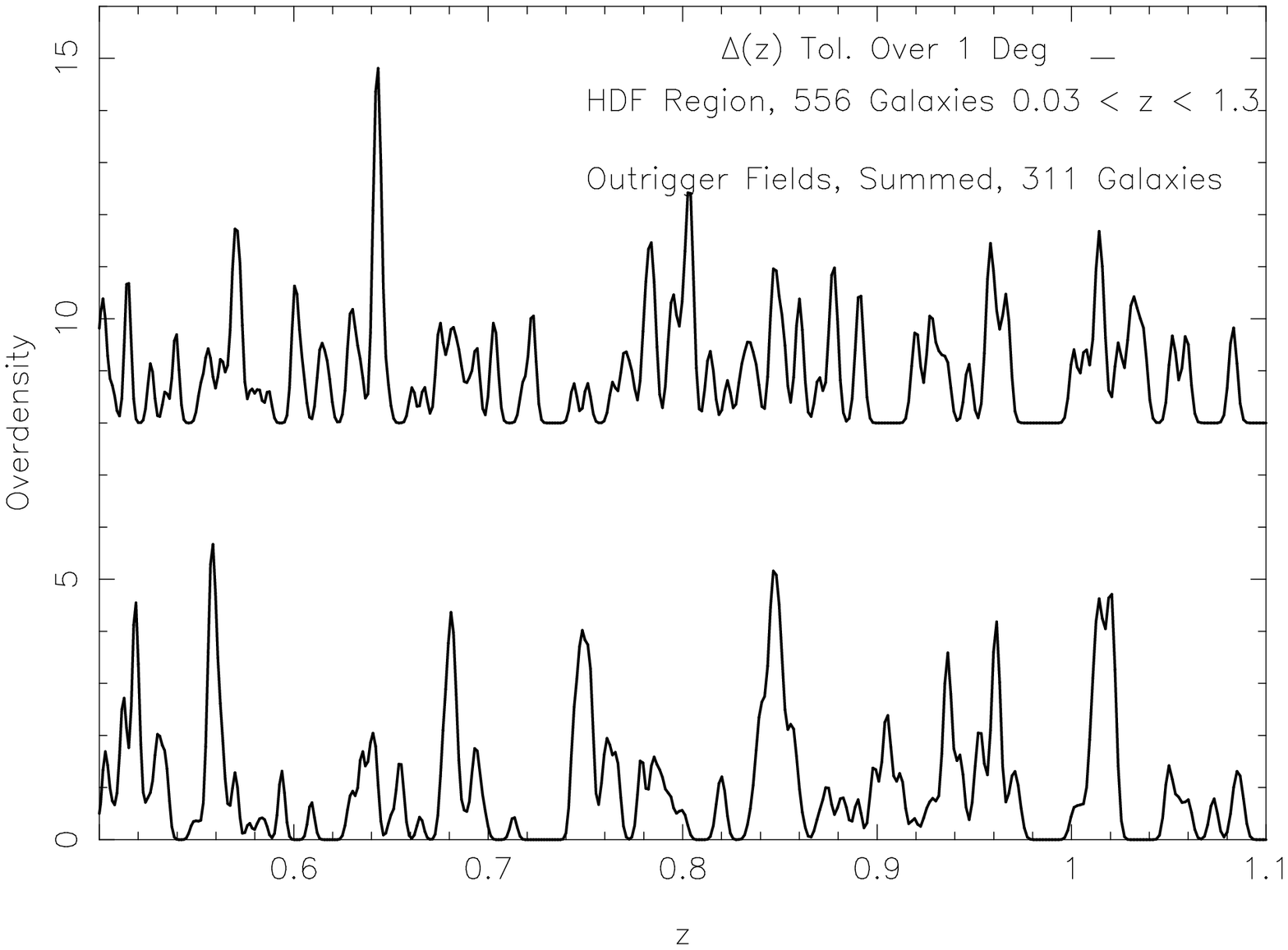} \\
\end{center}
{\sl Figure 4.  The overdensity group finding function is shown for the HDF
for $0.5 < z < 1.1$.
The overdensity function for four outrigger fields there taken together
is shown shifted upward.  The short horizontal line denotes
the expected tolerance in matching peaks across fields 1 deg apart.}

\section{Summary}

Our survey demonstrates very strongly that galaxies are clustered
on a small scale at all redshifts up to at least $z \sim 1.2$,
and that these structures are very similar in many respects
such as size, velocity dispersion, and total luminosity 
to groups and small clusters of galaxies as seen locally.

In terms of large scale structure, our survey suggests a characteristic
scale near 70 Mpc with (H$_0$ = 60 \kmsM). 
The physical processes that could lead to
such a scale being imprinted on the fluctuation spectrum are
discussed in Szalay (1999).
The characteristic scale we have found is significantly
smaller that the deduced locally from the LCRS (Shectman \etal\ 1996)
of 128 Mpc
by Doroshkevich \etal\ (1996) and of 215 Mpc by Broadhurst
\etal\ (1990).  We do not find the strict periodicity claimed
by Broadhurst \etal.

We offer up one final comment.
Because of the ubiquitous presence of groups of galaxies, and
the relatively small $z$ separation between such, it is going to be quite
difficult to use photometric redshift techniques to work on galaxy
clustering problems.  Photo-$z$s can be accurate to perhaps 10\%,
adequate for galaxy luminosity functions, but
that is not sufficient here.  This problem is going to have to be resolved
the old fashioned way, with real spectroscopic redshifts, big surveys,
and large samples.  Its going to take a while, even with the current
generation of 8 - 10 m telescopes.

\acknowledgments

I thank my collaborators Roger Blandford of Caltech and David Hogg of the
Institute for Advanced Study, with whom most of this work was done.
I am grateful for partial support from STScI/NASA grant AR-06337.12-94A.

\end{document}